\documentclass[11pt,a4paper]{article}
\usepackage{authblk}
\usepackage[hyperref]{emnlp2020}
\usepackage{times}
\usepackage{url}
\usepackage{latexsym}
\usepackage{enumitem}
\usepackage{color, colortbl}

\usepackage[hang,flushmargin]{footmisc}

\usepackage{microtype}
\aclfinalcopy

\usepackage{booktabs}
\usepackage{multirow}
\usepackage{adjustbox}
\usepackage{arydshln}
\usepackage{tabularx}
\usepackage{array}
\usepackage{graphicx,subcaption}
\usepackage{enumitem}
\usepackage{amsmath}
\usepackage{amsfonts}
\usepackage{amssymb}
\usepackage{dsfont}

\definecolor{bluencs}{rgb}{0.0, 0.53, 0.74}

\definecolor{Gray}{gray}{0.9}

\let\oldFootnote\footnote
\newcommand\nextToken\relax

\renewcommand\footnote[1]{%
    \oldFootnote{#1}\futurelet\nextToken\isFootnote}

\newcommand\isFootnote{%
    \ifx\footnote\nextToken\textsuperscript{,}\fi}

\newenvironment{rcases}
    {\left\lbrace\begin{aligned}}
    {\end{aligned}\right\rbrace}

\title{Distilling Dense Representations for Ranking\\ using Tightly-Coupled Teachers}

\author{Sheng-Chieh Lin\thanks{\hspace{0.16cm}Contributed equally.}\hspace{0.14cm}}
\newcommand\CoAuthorMark{\footnotemark[\arabic{footnote}]} % get the current value
\author{Jheng-Hong Yang\protect\CoAuthorMark \hspace{0.14cm}}
\author{Jimmy Lin}
\affil{David R. Cheriton School of Computer Science\\University of Waterloo}

\date{}

\begin{document}
\maketitle
\begin{abstract}
We present an approach to ranking with dense representations that applies knowledge distillation to improve the recently proposed late-interaction ColBERT model.
Specifically, we distill the knowledge from ColBERT's expressive MaxSim operator for computing relevance scores into a simple dot product, thus enabling single-step ANN search.
Our key insight is that during distillation, tight coupling between the teacher model and the student model enables more flexible distillation strategies and yields better learned representations.
We empirically show that our approach improves query latency and greatly reduces the onerous storage requirements of ColBERT, while only making modest sacrifices in terms of effectiveness.
By combining our dense representations with sparse representations derived from document expansion, we are able to approach the effectiveness of a standard cross-encoder reranker using BERT that is orders of magnitude slower.
\end{abstract}

\section{Introduction}
\label{sec:intro}

For well over half a century, solutions to the \textit{ad hoc} retrieval problem---where the system's task is return a list of top $k$ texts from an arbitrarily large corpus $\mathcal{C}$ that maximizes some metric of quality such as average precision or nDCG---has been dominated by {\it sparse} vector representations, for example, bag-of-words BM25.
Even in modern multi-stage ranking architectures, which take advantage of large pretrained transformers such as BERT~\cite{devlin2018bert}, the models are deployed as {\it rerankers} over initial candidates retrieved based on sparse vector representations; this is sometimes called ``first-stage retrieval''.
One well-known example of this design is the BERT-based reranker of~\citet{marco_BERT}.

The standard reranker architecture, while effective, exhibits high query latency, on the order of seconds per query~\cite{Hofstatter_Hanbury_2019,colbert} because expensive neural inference must be applied at query time on query--document pairs.
This design is known as a cross-encoder~\cite{Humeau_etal_ICLR2020}, and it exploits query--document attention interactions across all transformer layers.
As an alternative, the field has seen much recent interest in approaches based on representation learning that allow document representations to be precomputed independently of queries and stored.
Efficient libraries then allow large-scale comparisons between query and document vectors.
Overall, such approaches are less effective than cross-encoder reranking models, but far more efficient.

Within this general framework, we describe our low latency end-to-end approach for the \textit{ad hoc} passage retrieval task that combines dense and sparse representations.
As a starting point, we adopt the ``late interaction'' ColBERT model~\cite{colbert} and, via knowledge distillation~\cite{distilling_hinton}, are able to simplify its MaxSim relevance computation into dot-product similarity over pooled embeddings.
Since lexical signals (e.g., term frequencies) from sparse representations remain essential for \textit{ad hoc} retrieval~\cite{dpr, hybrid}, we further demonstrate that our dense representations can simply incorporate sparse signals without a complex joint training strategy~\cite{clear}.
In sum, we introduce simple-yet-effective strategies that leverage both dense and sparse representations for the end-to-end \textit{ad hoc} passage retrieval task.

\begin{figure}[t]
    \centering
    \resizebox{\columnwidth}{!}{
        \includegraphics{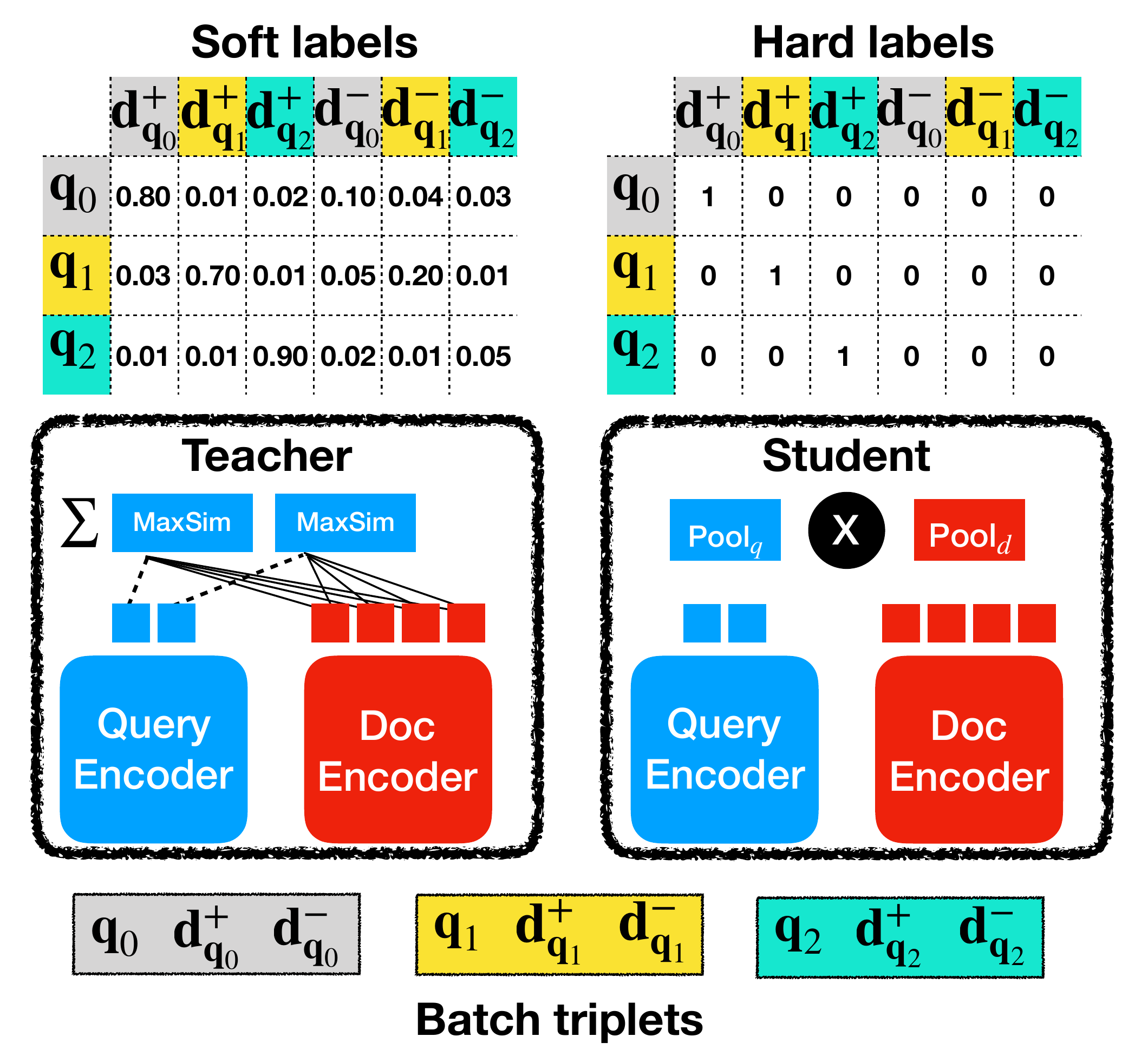}
    }
    \caption{Tight coupling between teacher and student models during distillation of dense representations for ranking.}
    \label{fig:idea}
\end{figure}

Our key insight is that during distillation, tight coupling between the teacher model and the student model enables more flexible distillation strategies and yields better learned representations (illustrated in Figure~\ref{fig:idea}).
By tight coupling, we mean that inference using the teacher model is interleaved directly during the distillation process:\ This is a key difference between our approach and previous methods where query--document scores are precomputed~\cite{Hofstatter:2010.02666:2020}.
With this tight coupling, we can also avoid computationally expensive mechanisms such as periodic index refreshes that are necessary during representation learning~\cite{guu2020realm,xiong2020approximate}.
A practical consequence of this tight coupling is that the teacher model must itself be reasonably efficient (thus, for example, ruling out teacher models based on cross-encoders).
For this role, ColBERT~\cite{colbert} is a good fit.

\section{Background}
\label{sec:dsp}

We begin by formalizing the representation learning problem for text ranking and review learning approaches.
We represent matrices by uppercase letters $X$, scalars by lowercase italic letters $x$, and vectors by lowercase bold letters $\mathbf{x}$.

The \textit{ad hoc} retrieval task can be viewed as a text ranking problem; here, we adopt the formulation of~\citet{lin2020pretrained}.
Specifically, we aim to learn some transformation $\eta_{(\cdot)}$, called an encoder, that maximize the following probability via surrogate functions given a pair comprising a query $\mathbf{q} \in \mathbb{R}^{n}$ and a candidate text (e.g., a passage) $\mathbf{d} \in \mathbb{R}^{n}$:
\begin{equation}
\label{eq:textrank}
\small
    P(\text{Relevant}\lvert\mathbf{q}, \mathbf{d}) \triangleq \phi(\eta_q(\mathbf{q}), \eta_d(\mathbf{d})),
\end{equation}
where $\phi$ is a similarity function and $n$ is an arbitrary natural number.
The system's task is to return the top-$k$ relevant texts for a query via the similarity function $\phi$ that takes $\eta_{q}(\mathbf{q})$ and $\eta_{d}(\mathbf{d})$.
Depending on the design, $\eta_q$ and $\eta_d$ can be identical or distinct.

\smallskip \noindent
\textbf{Dot-product similarity.}
Since online serving latency is critical in real-world applications, a standard choice of $\phi$ in Eq.~(\ref{eq:textrank}) is the dot product, $\langle\cdot, \cdot\rangle$.
Given this formulation, finding the top scoring passages that maximize $\langle\mathbf{q}, \mathbf{d}\rangle$ can be approximated efficiently by Approximated Nearest Neighbor search (ANN)~\cite{liu2004ann} or Maximum Inner Product Search (MIPS)~\cite{shrivastava2014mips} (henceforth, ANN), and accomplished by existing off-the-shelf libraries.

\smallskip \noindent
\textbf{Transformer-based bi-encoders.}
For large-scale applications, encoders based on pretrained transformers~\cite{devlin2018bert} have been widely adopted to map queries and passages into low dimensional vectors independently~\cite{lee-etal-2019-latent,Chang2020Pre-training,colbert,guu2020realm,dpr,hybrid,xiong2020approximate}.
Known as a bi-encoder design, a query (or a passage) is first mapped to a contextualized representation $E_{q} \in \mathbb{R}^{l_q \times t}$ (or $E_{d} \in \mathbb{R}^{l_d \times t}$), where $l$ indicates the length of the tokenized query (or the passage) of $t$-dimensional vectors.
Given the contextualized representation matrix, there are many choices for $\eta \colon \mathbb{R}^{l \times t} \to \mathbb{R}^{h}$ that transforms $E_{(\cdot)}$ into a lower dimensional vector $\eta(E_{(\cdot)})$, where $h \ll l \cdot t$.

Prior to retrieval, the $h$-dimensional representations can be precomputed for each of $|\mathcal{C}|$ texts in a corpus.
With specialized ANN libraries that take advantage of the parallelism provided by GPUs, even a brute force scan over millions of vectors is feasible.
With index structures, for example, based on small world graphs~\cite{malkov2020hnsw}, the ANN search problem can be further accelerated.

\smallskip \noindent
\textbf{Design choices.}
In general, compositions of $\phi$ and $\eta$ can be designed with different approaches.
For example, given a query--passage pair, we can define relevance in terms of the dot product between the two pooled embeddings as follows:
\begin{equation}
\small
\label{eq:pool}
    \phi_\text{PoolDot}(\mathbf{q},\mathbf{d}) = \langle \text{Pool}(E_{q}), \text{Pool}(E_{d}) \rangle ,
\end{equation}
where the Pool operator can be average or maximum pooling over token embeddings, or an indicator to a specific token embedding, e.g., the [CLS] embedding in BERT.
In this study, we adopt average pooling over token embeddings as our baseline, denoted as PoolAvg.

Our work builds on ColBERT~\cite{colbert}, who proposed a $\phi$ comparison function defined in terms of MaxSim, as follows:
\begin{equation}
\small
\label{eq:maxsim}
    \phi_\text{MaxSim}(\mathbf{q}, \mathbf{d}) = \sum_{i \in |E_q|} \max_{j \in |E_d|} \langle \eta_{q}({E_q}_{i}), \eta_{d}({E_d}_{j}) \rangle \text{,}
\end{equation}
where $\eta$ is composition of functions:
\begin{equation}
\small
\label{eq:colbert}
\begin{aligned}
    \eta_{q}(\textbf{x}) = \text{Normalize}(\text{Conv1D}(\textbf{x}))\\
    \eta_{d}(\textbf{x}) = \text{Filter}(\text{Normalize}(\text{Conv1D}(\textbf{x})).
\end{aligned}
\end{equation}
\noindent We refer readers to~\citet{colbert} for more details.
While ColBERT represents a design that greatly reduces retrieval latency with only a modest degradation in quality compared to the cross-encoder design, it still has two major limitations:

\begin{itemize}[leftmargin=*]
\item The process of computing Eq.~(\ref{eq:maxsim}) is approximated by a two-stage pipeline:\ retrieving then reranking, since MaxSim over the entire collection is not feasible.
Thus, despite its aspirations to single-stage ANN search, end-to-end retrieval with ColBERT still requires multi-stage retrieval.

\item The technique suffers from unreasonably high storage requirements compared to $\phi_\text{PoolDot}$ because the passages are preprocessed and stored as {\it sequences} of token embeddings via $\text{Conv1D}(E_{d}) \in \mathbb{R}^{l^{*} \times h^{*}}$, where $l^{*}$ denotes the length of the passage in tokens and $h^{*}$ denotes the kernel dimension of Conv1D in Eq.~(\ref{eq:colbert}).\footnote{\citet{colbert} append specialized tokens, [Q] and [D], for both queries and passages and set the kernel dimension of Conv1D to 128.}
\end{itemize}

\noindent On the other hand, learning well-behaved representations for the pooled embeddings using dot products directly, as in Eq.~(\ref{eq:pool}) is not trivial, since this process involves drastically non-linear dimension reduction.
The recent work of~\citet{guu2020realm} and~\citet{xiong2020approximate} propose adapting ANN search for mining hard negative examples to fine-tune the pretrained representations $E_{(\cdot)}$, which reduces the gap between training and inference.
However, this process is computationally demanding since it requires periodically refreshing the ANN index of all candidates (i.e., requiring inference over {\it all} texts in the corpus) to ensure that the best negative examples are retrieved.
\citet{xiong2020approximate} reports that re-encoding the entire corpus takes around 10 hours, and this occurs every 5K steps during training.

In another work, \citet{Hofstatter:2010.02666:2020} demonstrate that knowledge distillation from precomputed relevance scores of well-behaved cross-encoder rerankers is effective.
While distillation is able to capture reranking effectiveness, computationally expensive cross-encoder teachers limit the flexibility of exploring different combinations of query--document pairs, as exhaustively precomputing relevance scores using these cross-encoders can be computationally intractable.

\section{Methodology}

In contrast to the methods discussed above, we propose a  simple-yet-effective approach:\ knowledge distillation~\cite{distilling_hinton} with the novel insight that teacher and student models should be tightly coupled.
During training, in addition to fine-tuning using the contextualized representations $E_{(\cdot)}$ with relevance labels, we distill knowledge from ColBERT's similarity function $\phi_\text{MaxSim}$ into a dot-product bi-encoder.

Although ColBERT has enabled efficient passage retrieval, we seek to simplify it further.
To reduce computation and storage cost, we remove Conv1D and define our own similarity function in terms of average pooling over token embeddings (PoolAvg).
Thus, we precompute and store passage embeddings as $\text{Pool}(E_{d}) \in \mathbb{R}^{h}$ using ANN indexing in advance;
during inference, we only have to encode query embeddings as $\text{Pool}(E_{q}) \in \mathbb{R}^{h}$ and then conduct ANN search.\footnote{We set $h = 768$ for both queries and passages.}

\subsection{Knowledge Distillation}
\label{subsec:training}

Formally, given a query $\mathbf{q}$, we first estimate the relevance of a passage $\mathbf{d}$ using two sets of conditional probabilities:
\begin{equation}
\small
\begin{aligned}
   P(\mathbf{d}|\mathbf{q})= 
  \frac{\text{exp} (\phi_\text{PoolDot}(\mathbf{q}, \mathbf{d}))}{\sum_{\mathbf{d'}\in D} \text{exp} (\phi_\text{PoolDot}(\mathbf{q}, \mathbf{d'}))} \\
    \hat{P}(\mathbf{d}|\mathbf{q})= 
  \frac{\text{exp} (\phi_\text{MaxSim}(\mathbf{q}, \mathbf{d})/ \tau)}{\sum_{\mathbf{d'}\in D} \text{exp} (\phi_\text{MaxSim}(\mathbf{q}, \mathbf{d'})/ \tau)} & ,
   \end{aligned}
\end{equation}
where $\mathcal{D}$ is the set of all the passages, $\hat{P}$ is the relevance probability estimated by the knowledge source, and $\tau$ is the temperature to control the probability distribution.

Note that it is infeasible to enumerate all the passages during each training step; hence, following~\citet{Chang2020Pre-training}, we replace $\mathcal{D}$ with a sampled passage set $\mathcal{D}_\mathcal{B}$ in the same batch $\mathcal{B}$. 
Specifically, we have a batch of triplets $(\mathbf{q}_i, \mathbf{d}_{q_i}^{+}, \mathbf{d}_{q_i}^{-})_{i \in \mathcal{B}}$ as follows. For a query $\mathbf{q}_{i}$, we have: 

\begin{enumerate}[leftmargin=*]
\item a positive passage $\mathbf{d}_{q_i}^{+}$ in a positive labeled set $\mathcal{T}_{q_i}^{+}$, 
\item a negative passage $\mathbf{d}_{q_i}^{-}$ in a negative set $\mathcal{T}_{q_i;\text{BM25}}^{-}$ sampled by BM25 but not in $\mathcal{T}_{q_i}^{+}$, and 
\item the rest of passages for other queries $\{\mathbf{q}_j\}_{j\in\mathcal{B}}$ in the same batch: $\{\mathbf{d}_{q_j}^{+}\}_{j\in\mathcal{B}} \cup \{\mathbf{d}_{q_j}^{-}\}_{j\in\mathcal{B}}$.
\end{enumerate}
We denote the negative passage set $\mathcal{T}_{q_i;\mathcal{B}}^{-}$ for a query $q_i$ as the union of (2) and (3).
We train our model using the following objective function: 
\begin{equation}
\small
\begin{aligned}
\mathcal{L} = - \sum_{i=1}^{|\mathcal{B}|}
\begin{rcases}
\gamma \cdot \mathds{1}_{\mathbf{d}_i \in \mathcal{T}_{q_i}^{+}} \log(P(\mathbf{d}_i|\mathbf{q}_i)) - \\
(1-\gamma)\sum_{\mathbf{d'} \in \mathcal{D}_\mathcal{B}} \text{KL}( \hat{P}(\mathbf{d'}|\mathbf{q}_i) \lvert\rvert  P(\mathbf{d'}|\mathbf{q}_i)) \end{rcases} ,
\end{aligned}
\end{equation}
where the first term corresponds to the softmax cross entropy over relevance labels, the second term denotes the KL divergence between the sampled probability distributions from our teacher, the \textbf{T}ightly-\textbf{C}oupled \textbf{T}eacher ColBERT (TCT-ColBERT), and our student, the Siamese Network~\cite{bromley1993siamese} with BERT-base as encoders, denoted as bi-encoder (TCT-ColBERT).
The hyperparameter $\gamma$ controls the loss from hard and soft labels.\footnote{The pretrained weights for BERT-base are from \url{https://storage.googleapis.com/bert_models/2018_10_18/uncased_L-12_H-768_A-12.zip}.}
During the fine-tuning of the bi-encoder (TCT-ColBERT), we freeze the weight of ColBERT and set the temperature $\tau$ and $\gamma$ to 0.25 and 0.1, respectively.

\subsection{Hybrid Dense-Sparse Ranking}

As shown in~\citet{hybrid, clear}, a single dense embedding cannot sufficiently represent passages, especially when the passages are long, and they further demonstrate that sparse retrieval can complement dense retrieval by a linear combination of their scores.
However, it is not practical to compute scores over all query and passage pairs, especially when the corpus is large. 
Thus, we propose an alternative approximation, which is easy to implement.
In this work, we conduct end-to-end sparse and dense retrieval using Anserini~\cite{Yang_etal_JDIQ2018}\footnote{\url{https://github.com/castorini/anserini}} and Faiss~\cite{Johnson:1702.08734:2017},\footnote{\url{https://github.com/facebookresearch/faiss}} respectively.

For each query $\mathbf{q}$, we use sparse and dense representations to retrieve top 1000 passages, $\mathcal{D}_{sp}$ and $\mathcal{D}_{ds}$, with their relevance scores, $\phi_{sp}(\mathbf{q}, \mathbf{d}\in \mathcal{D}_{sp})$ and $\phi_{ds}(\mathbf{q}, \mathbf{d}\in \mathcal{D}_{ds})$, respectively.
Then, we compute the scores for each retrieved passages, $\mathbf{d} \in \mathcal{D}_{sp}\cup \mathcal{D}_{ds}$, as follows:

\begin{equation}
\label{eq:combination}
\small
\phi(\mathbf{q},\mathbf{d})=
\begin{cases} 
    \alpha \cdot \phi_{sp}(\mathbf{q},\mathbf{d}) +  \underset{\mathbf{d} \in \mathcal{D}_{ds}}\min_{}\phi_{ds}(\mathbf{q},\mathbf{d}), &\text{if }\mathbf{d} \notin D_{ds} \\
     \alpha \cdot \underset{\mathbf{d} \in \mathcal{D}_{sp}}\min_{}\phi_{sp}(\mathbf{q},\mathbf{d}) + \phi_{ds}(\mathbf{q},\mathbf{d}), & \text{if }\mathbf{d} \notin D_{sp} \\
      \alpha \cdot \phi_{sp}(\mathbf{q},\mathbf{d}) + \phi_{ds}(\mathbf{q},\mathbf{d}), & \text{otherwise.}\\
\end{cases}
\end{equation}
Eq.~(\ref{eq:combination}) is an approximation of linear combination of sparse and dense relevant scores. 
For approximation, if $\mathbf{d} \notin \mathcal{D}_{sp} (\text{or } \mathcal{D}_{ds})$, we directly use the minimum score of $\phi_{sp}(\mathbf{q}, \mathbf{d}\in \mathcal{D}_{sp})$, \text{or }$\phi_{ds}(\mathbf{q}, \mathbf{d}\in \mathcal{D}_{ds})$ as a substitute.

\begin{table*}[t]
	\caption{Main results on passage retrieval tasks.}
	\label{tb:main_result}
	\vspace{-0.2cm}
	\centering
	\small
    \begin{tabular}{lccccc}
	\toprule
	&\multicolumn{2}{c}{MS MARCO dev}& \multicolumn{2}{c}{TREC2019 DL}& latency\\
	\cmidrule(lr){2-3} \cmidrule(lr){4-5}
	&MRR@10& R@1000&NDCG@10& R@1000& (ms/query)\\
	\midrule
     \textbf{Sparse retrieval (Single Stage)}& &  &  &  &  \\
	   BM25 & 0.184 & 0.853 & 0.506 & 0.738 & 55 \\
	   DeepCT~\cite{deepct} & 0.243 & 0.913 & 0.551 & 0.756 & 55 \\
	   doc2query-T5~\cite{doctttttquery} & 0.277 & 0.947 & 0.642 & 0.802 & 64 \\
	\midrule
	 \textbf{Dense retrieval (Single Stage)}& &  &  &  &  \\
       ANCE~\cite{xiong2020approximate} & 0.330&0.959& 0.648 & - & 103\\
	   Bi-encoder (PoolAvg)& 0.310&0.945& 0.626 & 0.658 & 103\\
       Bi-encoder (TCT-ColBERT) & 0.335 & 0.964 & 0.670 & 0.720 & 103 \\
    \midrule
	 	 \textbf{Multi-Stage} & &  &  &  &  \\
        ColBERT~\cite{colbert} & 0.360 & 0.968&-& - & 458\\
	   BM25 + BERT-large~\cite{marco_BERT} &\textbf{0.365} &- &0.736 & -&3,500 \\
    \midrule
     \textbf{Hybrid dense + sparse (Single Stage)} & & & & & \\
       CLEAR~\cite{clear} &0.338 &0.969 & 0.699& 0.812&-\\
       Bi-encoder (PoolAvg) + BM25 &0.342 &0.962&0.701 &0.804& 106\\
       Bi-encoder (TCT-ColBERT) + BM25 &0.352 &0.970&0.714 &0.819& 106\\
       Bi-encoder (PoolAvg) + doc2query-T5 &0.354 &0.970 &0.719 &0.818 & 106\\
	   Bi-encoder (TCT-ColBERT) + doc2query-T5 &0.364 &\textbf{0.973} &\textbf{0.739} &\textbf{0.832} & 106\\
	\bottomrule
	\end{tabular}
	\vspace{0.2cm}
\end{table*}

\section{Experimental Setup}

To demonstrate the efficiency and effectiveness of our proposed design, we conduct experiments on a large-scale real world dataset.
We first describe the experiment settings and then elaborate on our empirical results in detail.

We conduct \textit{ad hoc} passage retrieval on the MS MARCO ranking dataset (henceforth, MS MARCO)~\cite{marco}.
It consists a collection of 8.8M passage from web pages and a set of 0.5M relevant (question, passage) pairs as training data, where each query on average has one relevant passage.
We follow two protocols for evaluation aligned with previous work~\cite{doctttttquery, deepct, clear, hybrid, colbert}:

\begin{itemize}[leftmargin=0.65cm]
\item[(a)] MS MARCO Dev: 6980 queries comprise the development set for MS MARCO, with on average one relevant passage per query.
We report MRR@10 and R@1000 as top-$k$ retrieval measures.
\item[(b)] TREC-2019 DL~\cite{trec19dl}:\ the organizers of the 2019 Deep Learning track at the Text REtrieval Conference (TREC) released 43 queries with multiple graded relevance labels, where 9k (query, passage) pairs were annotated by NIST assessors.
We report NDCG@10 and R@1000 for this evaluation set.
\end{itemize}

\noindent There are two steps in our training procedure:\ (1) fine-tune $\phi_ \text{MaxSim}$ as our teacher model, (2) freeze $\phi_ \text{MaxSim}$ and distill knowledge into our student model while fine-tuning $\phi_ \text{Pool}$. 
For both steps, we train models on the MS MARCO ``small'' triples training set for 160k iterations with a batch size of 96.
Note that at the second stage, we initialize the student model using the trained weights of the teacher model.
We fix sequence length to 32 and 150 for queries and passages, respectively.
For the sparse and dense retrieval combination, we tune the hyperparameter $\alpha$ on 6000 randomly sampled queries from the 0.5M queries with relevance labels for training (the ``train qrels'').
We conduct dense--sparse hybrid experiment with sparse signals from the original passages (denoted BM25) and docTTTTTquery~\cite{doctttttquery} (denoted doc2query-T5).
The optimal $\alpha$ for BM25 and doc2query-T5 are 0.10 and 0.24 respectively.

\section{Results}

Our main results are shown in Table~\ref{tb:main_result}, which reports effectiveness metrics as well as query latency.
We divide different comparison conditions into four categories:\ sparse retrieval, dense retrieval, multi-stage, dense--sparse hybrid. 

The cross-encoder reranker of~\citet{marco_BERT} provides a point of reference for multi-stage designs.
While it is effective, the model is also very slow.
In comparison, ColBERT is much faster, with only a small degradation in effectiveness.
However, it still relies on a two-stage retrieval design, and is about four times slower than other single-stage dense retrieval ANN search methods. 

As far as we are aware, ANCE~\cite{xiong2020approximate} is the current state of the art for single-stage dense retrieval, but as we have explained, its asynchronous training requires re-encoding and re-indexing the whole corpus during training.
Our proposed method, bi-encoder (TCT-ColBERT) slightly outperforms ANCE in terms of ranking accuracy and recall.
To highlight the effectiveness of our training strategy, we report the effectiveness of the bi-encoder design without distillation, denoted bi-encoder (PoolAvg), for a fair comparison.
A sizeable effectiveness increase from bi-encoder (PoolAvg) to bi-encoder (TCT-ColBERT) is observed in both tasks:\ $+0.025\ (+0.056)$ in MRR@10 and $+0.019\ (+0.062)$ in R@1000 for MS MARCO (TREC2019 DL).

When we further incorporate sparse signals, our proposed method beats the current state of the art in hybrid approaches, CLEAR~\cite{clear}, in both the MS MARCO and TREC 2019 DL tasks. 
Combined with BM25, our model already exhibits better retrieval effectiveness than CLEAR.
In addition, the comparison between bi-encoder (PoolAvg) and bi-encoder (TCT-ColBERT) demonstrates that the gain from distilled dense representation is still present, even with the advanced sparse retrieval method doc2query-T5:\ $+0.010\ (+0.010)$ and $+0.013\ (+0.020)$ with BM25 (doc2query-T5) in MRR@10 for both the MS MARCO and TREC2019 DL tasks, respectively.
The advanced hybrids (entries with doc2query-T5) reaches effectiveness even better than ColBERT and is almost on par with the cross-encoder reranker.
It is also worth noting that our hybrid end-to-end retrieval method yields state-of-the-art recall in both tasks.
More importantly, our proposed method is four times and thirty times more efficient than the multi-stage methods:\ ColBERT and the cross-encoder reranker, respectively.
These results demonstrate that the dense--sparse hybrid is a promising solution for low latency end-to-end text retrieval. 

\begin{table}[t]
	\caption{Component latency.}
	\label{tb:latency}
	\vspace{-0.2cm}
	\centering
	\small
    \begin{tabular}{lrc}
	\toprule
	Stage& latency & device\\
   & (ms/query) & \\
	\midrule
    BERT query encoder & 3 & GPU\\
    Dot product search & 100 & GPU\\ 
    Score combination & 3 & CPU\\
	\bottomrule
	\end{tabular}
	\vspace{0.2cm}	
\end{table}

\smallskip \noindent
\textbf{Latency.}
Table~\ref{tb:latency} shows the breakdown of end-to-end retrieval latency into individual components.
Specifically, we measure the system overhead of query embedding generation, dense retrieval with top 1000 passages, and dense--sparse score combination.
To obtain the latency for dense retrieval, we run BERT query encoder and dot product search using a 32GB V100 GPU.
Specifically, we conduct brute force dot product search in Faiss (indexing with IndexFlatIP).
As for the dense--sparse hybrid, we assume sparse and dense retrieval can be run in parallel; this is a realistic assumption because sparse retrieval runs on CPUs.
Thus, the total latency of the hybrid model (shown in Table~\ref{tb:main_result}) is bound by dense retrieval with additional 3ms for score combination (since sparse retrieval is faster than dense retrieval).

\smallskip \noindent
\textbf{Ablation study.}
Finally, we study the effectiveness of our distilled dense representations on the MS MARCO development set under two settings, reranking and retrieval.
For reranking, we use the public development set retrieved using BM25 for the reranking task (provided by the organizers),
and conduct reranking using dot product scores; for retrieval, we conduct brute force dot product search over the whole corpus.
We split our distillation strategy into two key features of our proposed technique:\ triplet and in-batch subsampling, from which we expect to see the effectiveness of triplet distillation (condition~2) and in-batch subsampling distillation (condition~3).
Specifically, by triplet distillation (condition~2) we mean that for each query $\mathbf{q}_i$, we only compute soft labels of its triplet ($\mathbf{q}_i, \mathbf{d}_{q_i}^{+}, \mathbf{d}_{q_i}^{-}$) for distillation instead of the whole in-batch samples (condition~3). 

\begin{table}[t]
	\caption{Ablation study on MS MARCO dev set.}
	\label{tb:ablation}
	\vspace{-0.2cm}
	\centering
	\small
	\resizebox{\columnwidth}{!}{
    \begin{tabular}{ccccc}
	\toprule
   &\multicolumn{2}{c}{Distillation strategy} & \multicolumn{2}{c}{MRR@10}\\
   \cmidrule(lr){2-3} \cmidrule(lr){4-5}
	Cond. & Triplet & In-batch & Re-ranking & Retrieval\\
	\midrule
    1&  & & 0.319 & 0.310\\
    \rowcolor{Gray}
    2&$\checkmark$ & & 0.332 & 0.328\\ 
    3&$\checkmark$ & $\checkmark$& 0.332 & 0.335\\
	\bottomrule
	\end{tabular}
	}
\end{table}

Table~\ref{tb:ablation} reports the ranking accuracy in terms of MRR@10.
First, we observe reranking yields better effectiveness than retrieval in conditions 1 and 2.
This indicates retrieval is a more challenging task than reranking, and potentially explains the discrepancy between training and inference noted by~\citet{xiong2020approximate}.
That is, in the training phase, the models only learn to discriminate positive passages from BM25-generated negative samples, which is similar to the reranking task; however, when conducting retrieval, models are required to rank documents from the whole corpus.
Despite the discrepancy between training and retrieval, in-batch subsampling (condition 3) shows better retrieval accuracy.
We attribute this to the distilled knowledge from in-batch samples.

Correspondingly, the superior effectiveness from in-batch subsampling showcases a key advantage of our design because the dynamic subsampling is feasible only when using a tightly-coupled teacher.
More advanced sampling methods such as importance sampling beyond uniform in-batch subsampling can be incorporated with our tightly-coupled teacher method, which we leave for future work.

\section{Conclusions}

Learned dense representations for ranking have recently attracted the attention of many researchers.
This approach is exciting because it has the potential to supplement, and perhaps even replace, sparse vector representations using inverted indexes.
There are no doubt many concurrent explorations along these lines, and we add our own contributions to the mix.
Knowledge distillation is a promising approach, and even beyond our specific approach built on ColBERT, we believe that our insight of tighter teacher--student coupling can be applied to other models and contexts as well.

\section*{Acknowledgements}

This research was supported in part by the Canada First Research Excellence Fund and the Natural Sciences and Engineering Research Council (NSERC) of Canada.
Additionally, we would like to thank Google for computational resources in the form of Google Cloud credits.

\bibliographystyle{acl_natbib}
\bibliography{paper.bib}

\end{document}